\documentclass[12pt]{article}
\usepackage{graphicx,epsfig}
\usepackage{osajnl2}
\usepackage{amsmath}
\begin{document}

\title{Theoretical study of optical fiber Raman polarizers with counterpropagating
beams}

\author{Victor V. Kozlov$^{1,2,*}$, Javier Nu$\bar{\hbox{n}}$o$^3$,
Juan Diego Ania-Casta$\tilde{\hbox{n}}$\'{o}n$^3$, and Stefan Wabnitz$^{1}$}
\address{$^1$Department of Information Engineering, Universit\`{a} di
Brescia, Via Branze 38, 25123 Brescia, Italy\\
$^2$Department of Physics, St.-Petersburg State University,
Petrodvoretz, St.-Petersburg, 198504, Russia\\
$^3$Instituto de Optica, Consejo Superior de Investigaciones
Cientificas (CSIC), 28006 Madrid, Spain}
\address{$^*$Corresponding author: victor.kozlov@email.com}
%\email{victor.kozlov@email.com}

\begin{abstract}
The theory of two counter-propagating polarized beams interacting in a randomly
birefringent fiber via the Kerr and Raman effects is developed and applied
to the quantitative description of Raman polarizers in the undepleted
regime. Here Raman polarizers, first reported by Martinelli {\it et. al.} 
[Opt. Express. {\bf 17}, 947 (2009)], 
are understood as Raman amplifiers
operating in the regime in which an initially weak unpolarized beam is
converted into an amplified fully polarized beam towards the fiber output.
Three parameters are selected for the characterization of a Raman
polarizer: the degree of polarization of the outcoming beam, its state of
polarization, and its gain. All of these parameters represent quantities that are averaged
over all random polarization states of the initially unpolarized
signal beam. The presented theory is computer friendly and applicable to virtually
all practically relevant situations, including the case of co-propagating beams,
and in particular to the undepleted as well
as the depleted regimes of the Raman polarizer.
\end{abstract}


\maketitle

%===============
\section{Introduction}
Fiber optic Raman amplifiers constitute an integral part of many
contemporary high-speed optical networks. Owing to their
broad amplification bandwidth, Raman amplifiers successfully compete with
erbium-doped-fiber amplifiers, \cite{review}. By the nature of the process
of stimulated Raman scattering (SRS) in silica fibers, Raman
amplifiers are highly polarization-dependent: the gain coefficient
reaches its maximum value whenever the state of polarization (SOP) of the
signal is parallel to the pump SOP. Whereas the amplifier gain is reduced by
two orders of magnitude if the two SOPs are orthogonal,
\cite{pd1,pd2,pd3}. In real fiber networks the fiber birefringence
changes stochastically along a fiber span. The net effect of this stochastically varying
birefringence is to greatly reduce the polarization dependence
of Raman gain. As a result, the effective Raman gain reduces to its span-averaged value, which is equal to
approximately half of its maximum value. Such averaging of the
gain between two orthogonal states of polarization is a
favorable feature, since polarization-dependent effects are
undesirable in most of the telecom applications which have been developed so far. The
value of the polarization-mode dispersion (PMD) coefficient
$D_p$ quantifies the degree of gain averaging, which
appears to be rather efficient for  $D_p>0.2$~ps$/\sqrt{\hbox{km}}$.
For smaller values of $D_p$, the onset of polarization-dependent
Raman gain is typically combated by scrambling the pump SOP.

In this paper our interest diverges from the common trend, as it
focuses on the exploitation of the polarization dependence of Raman 
gain. Motivated by the recent developments of transmission protocols 
based on polarization multiplexing \cite{pol_multiplex}, we are interested 
in developing nonlinear devices for achieving all-optical and ultrafast  
polarization control. Thus, we are interested in the use of fiber with low
values of the PMD coefficient, and do not perform the scrambling
of the pump beam. Moreover, the stronger the
polarization-dependent gain, the better the performance of
our device -- a Raman polarizer. These devices are Raman
amplifiers that are being fed by a weak unpolarized light signal,
and convert it into a powerful yet highly polarized light towards the
output. Preliminary theoretical investigations of Raman
polarizers and their proof-of-principle experiment, both in the
co-propagating geometry, have been recently reported in
Ref.~\cite{martinelli}. This geometry has been analyzed in
more details in the undepleted regime in Ref.~\cite{short}
%and in the depleted regime in Ref.~\cite{Luca}.

Here we set our goal in developing a simple and physically
transparent theory of Raman polarizers which is applicable to both
co-propagating and counter-propagating geometries, and
suitable for further use as a tool for analyzing the operation of these devices.
Our goal is also to address the two main issues which are related to
the performance of Raman polarizers. The first issue is
the degree of polarization (DOP) of the outcoming signal.
Given that the average DOP of the scrambled signal is zero initially, we say that our goal is
reached if the DOP of the output beam is close to unity. In
this case, we shall refer to such Raman amplifiers as ideal
Raman polarizers. The second issue is provided by the output signal SOP;
namely, its relation to the pump SOP. The DOP and SOP
of the outcoming signal fully characterize the performance of
Raman polarizers. The third important quantity is the value of
gain of the Raman polarizer. We shall comment on this issue,
too.

 It is important to distinguish from the outset between conventional Raman amplifiers
and Raman polarizers. Typical Raman amplifiers for telecom
applications are long, often surpassing $10$~km, require pump powers
of the order of $1$~W, and operate with relatively high values
of the PMD coefficient. In contrast, Raman polarizers are shorter, requiring a
fiber length of the order of $2$~km, pumped with high
power light -- of the order of $5$~W and higher, and favor
low values of the PMD coefficient. The last two characteristics
substantially strengthen the polarization dependence of the
gain, as we shall see from the subsequent analysis. Therefore the
operational regimes of Raman amplifiers and Raman polarizers
are drastically different, thus preventing the results of the theory
of Raman polarizers to become a trivial copy of the results already
known from the theory of Raman amplifiers.

The description of polarization-dependent gain in randomly
birefringent fibers is based on the vector theory of SRS, as it was
developed for instance in
Refs.~\cite{agrawal1,agrawal2,sergeev,galtarossa}. Theories
presented in Refs.~\cite{agrawal2,sergeev} bring analytical
insights into the problem, but their applicability is limited to the
regime in which the beat length of the fiber $L_B$ is substantially
smaller than the birefringence correlation length $L_c$:
$L_B\ll L_c$. Because of this restriction, those theories are not suitable
for the proper description of Raman polarizers, for which the opposite
inequality usually holds, see Ref.~\cite{short} and the analysis
that follows. In contrast, the theory in Ref.~\cite{galtarossa}
is most general and accurate, however it is resource-consuming
in terms of computational time, as it requires $10^3\div 10^4$
longer integration time than the theory which is presented here. Moreover,
since the theory of Ref.~\cite{galtarossa} is not formulated
in terms of deterministic differential equations, it is rather difficult
to extract from it physically transparent information about the role of the
different processes ruling the overall polarization dynamics.

As a matter of fact our theory is essentially the generalization of one-beam theory
of Wai and Menyuk in Ref.~\cite{wai_menyuk} to the case of two beams
interacting not only via the Kerr but also via the Raman effect.
It is close to the approach that was undertaken in Ref.~\cite{sergeev},
but it proceeds till the end with virtually no approximations. The
only important requirement here is that $\min (L,\, L_{NL})\gg L_c$,
where $L$ is the total length of the fiber and $L_{NL}$ its
nonlinear length. This inequality holds true for almost all
practically relevant situations. Another assumption in our theory
is that in the randomly varying birefringence tensor
characterizing the fiber
\begin{equation}
\Delta\boldsymbol{B} =\Delta\beta (\omega )
\left(\cos\theta \boldsymbol{\sigma}_3
+\sin\theta \boldsymbol{\sigma}_1\right)\, ,
\label{1}
\end{equation}
we shall only treat the orientation of the birefringence axis
$\theta$ as a stochastic variable, while we keep the value of the
birefringence $\Delta\beta$ as a constant (this is the so-called fixed modulus
model). It is now a well established fact that the theory with
$\theta$ as the only stochastic variable and the theory where
both $\theta$ and $\Delta\beta$ are stochastic variables
(random modulus model) both produce virtually identical results.
In Eq.~(\ref{1}), $\boldsymbol{\sigma}_1$ and
$\boldsymbol{\sigma}_3$ are usual Pauli matrices. Note that
we define the birefringence beat length as $L_B=2\pi /\Delta\beta$.

%===========
\section{Model}
Starting from the most basic model of interaction of two fields within a
Kerr- and Raman-active medium, whose tensorial response
as it is relevant to fused silica is properly taken into account, see
Ref.~\cite{agrawal2}, we follow the lines of derivations that were outlined in
Ref.~\cite{tutorial} and formulate the equation of motion of the
pump beam in the form
\begin{eqnarray}
&& \pm i\frac{\partial U_p}{\partial z}
+i\beta^\prime (\omega_p)\frac{\partial U_p}{\partial t}=
-\Delta\boldsymbol{B}U_p
-\frac{1}{3}\gamma_{pp}\left[2(U_p^*\cdot U_p)U_p
+(U_p\cdot U_p)U_p^*\right]
\nonumber\\
&& -\frac{2}{3}\gamma_{ps}
\left[(U_s^*\cdot U_s)U_p+(U_s\cdot U_p)U_s^*
+(U_p\cdot U_s^*)U_s\right]
+i\epsilon_p g_0(U_p\cdot U_s^*)U_s\, .
\label{2}
\end{eqnarray}
Here $U_p=(U_{px},\, U_{py})^T$ is the field vector written
in terms of normal polarization modes $e_x$ and $e_y$.
$\beta (\omega_p)$ is the propagation constant and
$\beta^\prime (\omega_p)$ is its frequency derivative.
The upper sign (``$+$") describes the configuration when
the signal and pump beams propagate in the fiber in one
direction (co-propagating geometry), while the lower sign
(``$-$") is related to the situation when they propagate in
opposite directions (counter-propagating geometry).
The self-polarization modulation (SPM) coefficient $\gamma_{pp}$
is the usual nonlinear coefficient of silica, $\gamma$, calculated
at the central frequency $\omega_p$ of the pump beam.
Cross-polarization modulation (XPM) coefficient
$\gamma_{ps}$ is also equal to $\gamma (\omega_p)$; $g_0$
is the Raman gain coefficient; $\epsilon_p=-\omega_p/\omega_s$.
The equations for the signal beam and all coefficients are the same
as above but with indices $p$ and $s$ interchanged, and with
$\epsilon_s=1$.

First, let us transform the equation (\ref{2}) to the local axes
of birefringence by performing the rotation of the field
vector $U_j$ as
\begin{equation}
\bar{U}_j=\left(
\begin{array}{cc}
\cos\frac{\theta}{2} & \sin\frac{\theta}{2} \\
-\sin\frac{\theta}{2} & \cos\frac{\theta}{2}
\end{array}\right)
U_j\, ,
\label{3}
\end{equation}
where $j=p,\, s$. Equation (\ref{2}) is not altered by this
transformation, the only difference being the change of the
form of the birefringence tensor, which now becomes
\begin{eqnarray}
&& \overline{\Delta \boldsymbol{B}}(\omega_p)=
\left(
\begin{array}{cc}
\Delta\beta (\omega_p) & \mp\frac{i}{2}\theta_z \\
\pm\frac{i}{2}\theta_z & -\Delta\beta (\omega_p)
\end{array}\right)\, ,
\label{4}\\
&& \overline{\Delta \boldsymbol{B}}(\omega_s)=
\left(
\begin{array}{cc}
\Delta\beta (\omega_s) & -\frac{i}{2}\theta_z \\
\frac{i}{2}\theta_z & -\Delta\beta (\omega_s)
\end{array}\right)\, .
\label{4_1}
\end{eqnarray}
Here, $\theta_z$ is the derivative of
$\theta$ with respect to $z$. It is different from zero
owing to the random changes of orientation of the 
birefringence axes. Namely, the change of $\theta$
is driven by the white noise process:
$\theta_z=g_\theta (z)$, with zero mean and
$\langle g_\theta (z)g_\theta (z^\prime )\rangle
=2L_c^{-1}\delta (z-z^\prime )$, where $L_c$ is the
birefringence correlation length, as mentioned in the
Introduction.

In line with our assumption that $\min (L,\, L_{NL})\gg L_c$,
we separate the fast motion related to the rapid changes
of $\theta$ from the slow motion that occurs on the scale 
of the nonlinear
length. It is important to note that this separation is exact
and yet does not involve any approximation. The
approximation is to be made at a later stage.
The required transformations read as
\begin{eqnarray}
&& V_p=\boldsymbol{T}_p(z)\bar{U}_p=\left(
\begin{array}{cc}
a_1 & a_2\\
-a_2^* & a_1^*
\end{array}\right)U_p\, ,
\label{5}\\
&& V_s=\boldsymbol{T}_s(z)\bar{U}_s=\left(
\begin{array}{cc}
b_1 & b_2\\
-b_2^* & b_1^*
\end{array}\right)U_s\, .
\label{6}
\end{eqnarray}
Here $V_p=(V_{p1},\, V_{p2})^T$ and
$V_s=(V_{s1},\, V_{s2})^T$.
Matrices $\boldsymbol{T}_p$ and $\boldsymbol{T}_s$
obey equations of motion
\begin{eqnarray}
\pm i\frac{\partial \boldsymbol{T}_p}{\partial z}
+\overline{\Delta\boldsymbol{B}}(\omega_p)
\boldsymbol{T}_p=0\, ,
\label{7}\\
i\frac{\partial \boldsymbol{T}_s}{\partial z}
+\overline{\Delta\boldsymbol{B}}(\omega_s)
\boldsymbol{T}_s=0\, ,
\label{7_1}
\end{eqnarray}

These transformations eliminate the birefringence
terms from the equations of motion of $V_p$ and $V_s$
and bring about a vast number of cubic terms composed
of different combinations of $V_{p1}$, $V_{p2}$,
$V_{s1}$, $V_{s2}$ and their complex conjugates.
Factors in front of these terms are products of two
coefficients of the form $u_mu_n$, or $u_m^*u_n$,
or $u_m^*u_n^*$, where $m,n=1,\,\dots ,\, 14$.
Products with $m=n$ we shall call self-products,
while with $m\ne n$ cross-products. Here,
$u_1=\vert a_1\vert^2-\vert a_2\vert^2$,
$u_2=-(a_1a_2+a_1^*a_2^*)$,
$u_3=i(a_1a_2-a_1^*a_2^*)$,
$u_4=2a_1a_2^*$, $u_5=a_1^2-{a_2^*}^2$,
$u_6=-i(a_1^2+{a_2^*}^2)$,
$u_7=a_1^*b_1-a_2b_2^*$,
$u_8=-(b_1a_2+b_2^*a_1^*)$,
$u_9=i(b_1a_2-a_1^*b_2^*)$,
$u_{10}=-i(a_1^*b_1+a_2b_2^*)$,
$u_{11}=a_1b_2^*+b_1a_2^*$,
$u_{12}=a_1b_1-a_2^*b_2^*$,
$u_{13}=-i(a_1b_1+a_2^*b_2^*)$,
$u_{14}=i(a_1b_2^*-a_2^*b_1)$.

In the thus obtained equations of motion for
$V_p$ and $V_s$ we perform the ensemble
average (over realizations of the random process that describes the linear birefringence). 
Thus, we write $\langle u_m u_n\rangle$
instead of $u_mu_n$. This change holds true
only in the limit when the stochastic variations are
faster than the nonlinear beam evolution. This is exactly
the place in the derivation where our single
approximation comes into play. At this point we
also need to apply the ergodic theorem
\begin{equation}
\langle f\rangle =
\lim_{z\to\infty}\frac{1}{z}\int_0^zdz^\prime\, f(z^\prime )\, .
\label{8}
\end{equation}
Our goal is to calculate ensemble averages of all
necessary self- and cross-products, and on this way we may
complete the derivation of the differential equations for
$V_p$ and $V_s$.

The equations of motion for $u_n$ with $n=1,\,\dots ,\, 14$
can be easily formulated basing ourselves on equations (\ref{7})
and (\ref{7_1}). As these equations are linear, in order to find an
ensemble average of any function of these coefficients (
in our case pair products) we need to construct a
generator. We refer to the Appendix in
Ref.~\cite{wai_menyuk} for details of this procedure,
and only give here the final result. With this generator
we are able to formulate the equations of motion for the
ensemble averages of the products of the coefficients.
Thus the solutions to the equations of motion
\begin{eqnarray}
&& \frac{\partial G_1}{\partial z}=
-\frac{2}{L_c}(G_1-G_2)\, ,
\label{9}\\
&& \frac{\partial G_2}{\partial z}=
\frac{2}{L_c}(G_1-G_2)\mp 4\Delta\beta (\omega_p)G_4 \, ,
\label{10}\\
&& \frac{\partial G_3}{\partial z}=
\pm 4\Delta\beta (\omega_p)G_4 \, ,
\label{11}\\
&& \frac{\partial G_4}{\partial z}=
-\frac{1}{L_c}G_4\pm 2\Delta\beta (\omega_p)(G_2-G_3)
\label{12}
\end{eqnarray}
yield the result for the self-products
$\{ \langle u_1^2\rangle ,\, \langle  u_2^2\rangle ,\, \langle u_3^2\rangle\}$,

\noindent
$\{ \langle \hbox{Re}^2(u_4)\rangle ,\, \langle\hbox{Re}^2(u_5)\rangle ,\,
\langle \hbox{Re}^2(u_6)\rangle\}$, 

\noindent
and $\{ \langle \hbox{Im}^2(u_4)\rangle ,\,
\langle\hbox{Im}^2(u_5)\rangle ,\, \langle \hbox{Im}^2(u_6)\rangle\}$, if we
associate them with $\{ G_1,\, G_2,\, G_3\}$ with initial conditions
given as $(1,\, 0,\, 0)$, $(0,\, 1,\, 0)$, and $(0,\, 0,\, 1)$, respectively.

The remaining self-products
$\{ \langle \hbox{Re}^2(u_7)\rangle ,\, \langle\hbox{Re}^2(u_8)\rangle ,\,
\langle \hbox{Re}^2(u_9),\, \langle\hbox{Re}^2(u_{10})\rangle\}$,

\noindent
$\{ \langle \hbox{Im}^2(u_7)\rangle ,\, \langle\hbox{Im}^2(u_8)\rangle ,\,
\langle \hbox{Im}^2(u_9),\, \langle\hbox{Im}^2(u_{10})\rangle\}$,

\noindent
$\{ \langle \hbox{Re}^2(u_{11})\rangle ,\, \langle\hbox{Re}^2(u_{12})\rangle ,\,
\langle \hbox{Re}^2(u_{13}),\, \langle\hbox{Re}^2(u_{14})\rangle\}$,

\noindent
and $\{ \langle \hbox{Im}^2(u_{11})\rangle ,\, \langle\hbox{Im}^2(u_{12})\rangle ,\,
\langle \hbox{Im}^2(u_{13}),\, \langle\hbox{Im}^2(u_{14})\rangle\}$,
can be found from the equations
\begin{eqnarray}
&& \frac{\partial G_1}{\partial z}=-\frac{2}{L_c}(G_1-G_2)
+2\left[\pm\Delta\beta (\omega_p)-\Delta\beta (\omega_s)\right]G_5\, ,
\label{13}\\
&& \frac{\partial G_2}{\partial z}=\frac{2}{L_c}(G_1-G_2)
-2\left[\pm\Delta\beta (\omega_p)+\Delta\beta (\omega_s)\right]G_6\, ,
\label{14}\\
&& \frac{\partial G_3}{\partial z}=
2\left[\pm\Delta\beta (\omega_p)+\Delta\beta (\omega_s)\right]G_6\, ,
\label{15}\\
&& \frac{\partial G_4}{\partial z}=
-2\left[\pm\Delta\beta (\omega_p)-\Delta\beta (\omega_s)\right]G_5\, ,
\label{16}\\
&& \frac{\partial G_5}{\partial z}=
\left[\pm\Delta\beta (\omega_p)-\Delta\beta (\omega_s)\right](G_4-G_1)
-\frac{1}{L_c}G_5\, ,
\label{17}\\
&& \frac{\partial G_6}{\partial z}=
\left[\pm\Delta\beta (\omega_p)+\Delta\beta (\omega_s)\right](G_2-G_3)
-\frac{1}{L_c}G_6\, ,
\label{18}
\end{eqnarray}
when we associate them with
$\{ G_1,\, G_2,\, G_3,\, G_4\}$, with initial conditions
as $(1,\, 0,\, 0,\, 0)$, $(0,\, 0,\, 0,\, 1)$, $(0,\, 1,\, 0,\, 0)$,
and $(0,\, 0,\, 1,\, 0)$, respectively.

In order to find the cross-products we constructed appropriate
generators and found that all the cross-products that are of
interest to us turn out to be equal to zero. Similarly, terms of the form
$\hbox{Re}(u_n)\hbox{Im}(u_n)$ also vanish. Thus, many
SPM, XPM, and Raman terms in the final equations of motion
disappear. The thus found equations of motion for the fields
are conveniently formulated in Stokes space. They read as
\begin{eqnarray}
\left(\pm\frac{\partial}{\partial z}
+\beta^\prime (\omega_p)\frac{\partial}{\partial t}\right)
S^{(p)} & = & \gamma (\omega_p)
\left(S^{(p)}\times \boldsymbol{J}_S^{(p)}(z) S^{(p)}
+S^{(p)}\times \boldsymbol{J}_X(z) S^{(s)}\right)
\nonumber\\
&& +\epsilon_pg_0\left(S_{0}^{(s)}J_{R0}S^{(p)}
+S_0^{(p)}\boldsymbol{J}_R(z)S^{(s)}
\right)\, ,
\label{19}\\
\left(\frac{\partial}{\partial z}
+\beta^\prime (\omega_s)\frac{\partial}{\partial t}\right)
S^{(s)} & = & \gamma (\omega_s)
\left(S^{(s)}\times \boldsymbol{J}_S^{(s)}(z) S^{(s)}
+S^{(s)}\times \boldsymbol{J}_X(z) S^{(p)}\right)
\nonumber\\
&& +g_0\left(S_{0}^{(p)}J_{R0}S^{(s)}
+S_0^{(s)}\boldsymbol{J}_R(z) S^{(p)}\right)\, .
\label{20}
\end{eqnarray}
Here $S^{(j)}$ is a three-component vector:
$S^{(j)}=(S_1^{(j)},\, S_2^{(j)},\, S_3^{j})$,
whose components are defined as
\begin{eqnarray}
&& S_1^{j}=V_{j1}V_{j2}^*+V_{j1}^*V_{j2}\, ,
\label{21}\\
&& S_2^{j}=i(V_{j1}^*V_{j2}-V_{j1}V_{j2}^*)\, ,
\label{22}\\
&& S_3^{j}=\vert V_{j1}\vert^2 -\vert V_{j2}\vert^2\, ,
\label{23}
\end{eqnarray}
where $j=p,\, s$. It is instructive to formulate
separate equations for the powers of the signal
and pump beams, which may be defined through the components
of the Stokes vectors as
$S_0^{(j)}=\left({S_1^{(j)}}^2+{S_2^{(j)}}^2+{S_3^{(j)}}^2\right)^{1/2}$:
\begin{eqnarray}
&& \left(\pm\frac{\partial}{\partial z}
+\beta^\prime (\omega_p)\frac{\partial}{\partial t}\right)
S_0^{(p)} =
\nonumber\\
&& \epsilon_pg_0\left(
J_{R0}S_0^{(s)}S_0^{(p)}
+J_{R1}(z)S_1^{(s)}S_1^{(p)}
+J_{R2}(z)S_2^{(s)}S_2^{(p)}
+J_{R3}(z)S_3^{(s)}S_3^{(p)}
\right)\, ,
\label{24}\\
&& \left(\frac{\partial}{\partial z}
+\beta^\prime (\omega_s)\frac{\partial}{\partial t}\right)
S_0^{(s)} =
\nonumber\\
&& g_0\left(J_{R0}S_0^{(s)}S_0^{(p)}
+J_{R1}(z)S_1^{(s)}S_1^{(p)}
+J_{R2}(z)S_2^{(s)}S_2^{(p)}
+J_{R3}(z)S_3^{(s)}S_3^{(p)}\right)\, ,
\label{25}
\end{eqnarray}

Matrices in equations (\ref{19}) and (\ref{20})
are all diagonal with elements

\noindent
$\boldsymbol{J}_R=\hbox{diag}(J_{R1},\, J_{R2},\, J_{R3})$,
$\boldsymbol{J}_X=\hbox{diag}(J_{X1},\, J_{X2},\, J_{X3})$,
$\boldsymbol{J}_S=\hbox{diag}(J_{S1},\, J_{S2},\, J_{S3})$.
These elements are different for the counter-propagating
and the co-propagating interaction geometries.

In order to complete our theory, we need to express
all elements in these matrices in terms of ensemble
averages of self-products:
\begin{eqnarray}
&& J_{R1}=\langle \hbox{Re}(u_{14}^2-u_{10}^2)\rangle\, ,
\label{26}\\
&& J_{R2}=-\langle \hbox{Re}(u_{14}^2+u_{10}^2)\rangle\, ,
\label{27}\\
&& J_{R3}=-\langle \vert u_{14}\vert^2-\vert u_{10}\vert^2\rangle\, ,
\label{28}\\
&& J_{X1}=\frac{2}{3}\langle \hbox{Re}(u_{10}^2
+u_{13}^2-u_9^2-u_{14}^2)\rangle\, ,
\label{29}\\
&& J_{X2}=\frac{2}{3}\langle \hbox{Re}(u_{10}^2
+u_{14}^2-u_9^2-u_{13}^2)\rangle\, ,
\label{30}\\
&& J_{X3}=\frac{2}{3}\langle \vert u_{9}\vert^2+\vert u_{14}\vert^2
-\vert u_{13}\vert^2-\vert u_{10}\vert^2\rangle\, ,
\label{31}\\
&& J_{S1}=\frac{1}{3}\langle \hbox{Re}(u_{6}^2)\rangle\, ,
\label{32}\\
&& J_{S2}=-\frac{1}{3}\langle \hbox{Re}(u_{6}^2)\rangle\, ,
\label{33}\\
&& J_{S3}=\frac{1}{3}\left[3\langle u_3^2\rangle -1\right]\, ,
\label{34}
\end{eqnarray}
and also $J_{R0}=\langle \vert u_{10}\vert^2+\vert u_{14}\vert^2\rangle$.

%========================
\section{Physical considerations}
Model equations (\ref{19}) and (\ref{20}) represent the central finding
of our paper. These equations describe vector Raman amplification under
rather general conditions. The SPM, XPM, and Raman matrices
contain $z$-dependent elements on their diagonals, whose
values are expressed in terms of the self-product coefficients in
Eqs.~(\ref{26})-(\ref{34}), while the coefficients themselves
are found as the solutions of equations (\ref{9})-(\ref{12}) and
(\ref{13})-(\ref{18}).

We are about to apply these model equations to the
problem of Raman polarizers. Before treating this specific
case, some comment about the Raman matrix is in
order. It is commonly accepted that the net effect of
polarization-dependent gain is the attraction of the signal
SOP to the pump SOP. While this is certainly the case for
isotropic fibers, an extension of this idea to the case of
randomly birefringent fibers should be undertaken with
caution. As a matter of fact, we found that in general in randomly birefringent fibers the
signal SOP is not attracted to the pump SOP. In order
to demonstrate this, the most straightforward way is to
analyze equation (\ref{25}) for the power of the signal
beam. For isotropic fibers, we get $J_{R1}=J_{R2}=J_{R3}=1$,
so that parallel signal and pump SOPs provide the maximally
possible gain.

%=== Figure 1 ===
\begin{figure}
\begin{center}
\includegraphics[scale=0.6]{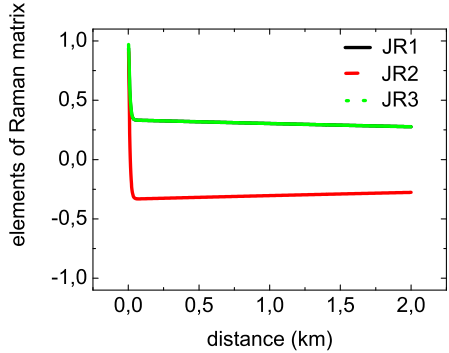}
\end{center}
\caption{Elements of the Raman matrix
$J_R=\hbox{diag}\left(J_{R1},\, J_{R2},\, J_{R3}\right)$
($J_{R1}$ -- black solid, $J_{R2}$ -- red dashed, and
$J_{R3}$ -- green dotted) as function of distance in
the fiber for $L_B(\omega_p)=50$~m and $L_c=1$~m.
(note that the black solid and green dotted curves
nearly coincide, i.e. $J_{R1}=J_{R3}$.}
\label{ris1}
\end{figure}
%=============

When we turn to the case of randomly birefringent fibers,
we find that the elements of the Raman matrix are not
always equal to each other in the co-propagating geometry,
as it was demonstrated in Ref.~\cite{short}. Moreover, these elements are never equal to each
other in the counter-propagating geometry, as it is exemplified
in Fig.~\ref{ris1}. As can be seen in this figure, the Raman matrix elements are not only different from
each other but they can also be negative.

Although in general the signal SOP is not attracted to the pump
SOP, the analysis of the co-propagating configuration in
Ref.~\cite{short} shows that in the regime when Raman
amplifiers act as ideal Raman polarizers the output signal SOP
is always almost perfectly aligned with the pump SOP. For this
regime, the polarization properties of the randomly birefringent
fibers are virtually the same as in the case of isotropic fibers.

On the other hand in the counter-propagating geometry the situation
is totally different. In the regime when Raman amplifiers act
as ideal Raman polarizers, the typical relation between the
diagonal elements of the Raman matrix is as shown in
Fig.~\ref{ris1}. Thus, the first and the third components of the signal
Stokes vector are attracted to the corresponding components
of the pump Stokes vector, while the second components are
repelled from each other. Therefore whenever the pump SOP
contains an appreciable admixture of the circular polarization,
the signal SOP is never attracted to the pump SOP. In the next
section we suggest a simple rule with a wide range of applicability
on how to determine the output signal SOP.

The polarization-dependent gain is quantified by the values
of the diagonal elements of the Raman matrix. The larger these
values, the better the performance of the Raman polarizers. In fibers
with high PMD values the diagonal elements quickly vanish
near the input fiber end already. For smaller PMD values the
$z$-dependent elements of the Raman matrix keep appreciable
values across the whole fiber span, as demonstrated in
Fig.~\ref{ris1}. Indeed, low PMD fibers are good candidates for the
implementation of Raman polarizers.

The elements of the XPM matrix exhibit a similar dependence upon the
PMD coefficient and have a magnitude which is comparable to that of the diagonal elements
of the Raman matrix. Nevertheless the nonlinear
polarization rotation which is due to the XPM interaction is very weak,
and it has virtually no effect on the performance of Raman polarizers operating
in the undepleted regime.

The SPM effect for the signal beam has also no impact on the
performance of Raman polarizers. First, this is because the diagonal
elements of the SPM matrix $\boldsymbol{J_S}^{(s)}$
vanish on first $100$~m of the fiber for the practically relevant
range of beat lengths and correlation lengths, a range that
we define as follows:
\begin{equation}
0.001<L_B<0.1\quad\hbox{and}\quad 0.0001<L_c<0.05\, ,
\label{35}
\end{equation}
which is given here in km. Second, because the signal beam is too
weak to experience a significant nonlinear self-interaction, especially near
its input end.

In contrast, the SPM effect can in principle be sizable for the
pump beam. Given that the pump power is relatively high
($8$~W in our simulations), even the first $100$~m are enough
to perturb the pump SOP. However, these perturbations remain relatively small
(of the order of $0.1$\%) for a pump power as high as $8$~W.

%===================
\section{Numerical results}
We performed extensive simulations in a configuration where
the pump and signal beams counter-propagate
through a $2$~km long fiber, and the pump power
is set to $8$~W at $z=L$. The input power of the
signal beam is set to be equal to $0.1$~mW, which is low enough
to ensure that Raman amplification occurs in the
undepleted regime for all $z$ and the entire range of fiber parameters. 
We varied the beat length and the correlation
length within the range that is defined in Eq.~(\ref{35}).
We followed two quantities of interest at the output
end of the fiber at $z=L$ as function of the beat and the
correlation lengths. The first such quantity is the DOP ($D$) of
the signal beam, which is the central characteristic
of Raman polarizers. Starting with an unpolarized
signal beam (with $D=0$), we aim at finding such
parameter regimes for which the outcoming signal
beam has DOP close to unity. The unpolarized
signal beam is modeled in the Stokes space as an
ensemble of beams with $N\sim 5000$ states of
polarizations that are uniformly distributed over the entire
Poincar\'{e} sphere. On the other hand the pump is a polarized beam
whose SOP is varied as detailed below. The
second quantity of interest is the so-called
alignment parameter, which is defined as
\begin{equation}
A_{\uparrow\downarrow}\equiv
\frac{\left\langle S_1^{(s)}S_1^{(p)}-S_2^{(s)}S_2^{(p)}
+S_3^{(s)}S_3^{(p)}\right\rangle}{S_0^{(s)}S_0^{(p)}}\, ,
\label{36}
\end{equation}
where the angle brackets indicate average over all
$N$ realizations of the SOPs of the input scrambled
signal beam. This quantity characterizes the average
orientation of the signal SOP with respect to the
pump SOP. Note that in the copropagating geometry the appropriate alignment
parameter reads as
\cite{short},
\begin{equation}
A_{\uparrow\uparrow}\equiv
\frac{\left\langle S_1^{(s)}S_1^{(p)}+S_2^{(s)}S_2^{(p)}
+S_3^{(s)}S_3^{(p)}\right\rangle}{S_0^{(s)}S_0^{(p)}}\, .
\label{36_1}
\end{equation}
This alignment parameter is nothing but an average
cosine of the angle between the pump and signal
SOPs in Stokes space.

%=== Figure 2 ===
\begin{figure}
\begin{center}
\includegraphics[scale=0.8]{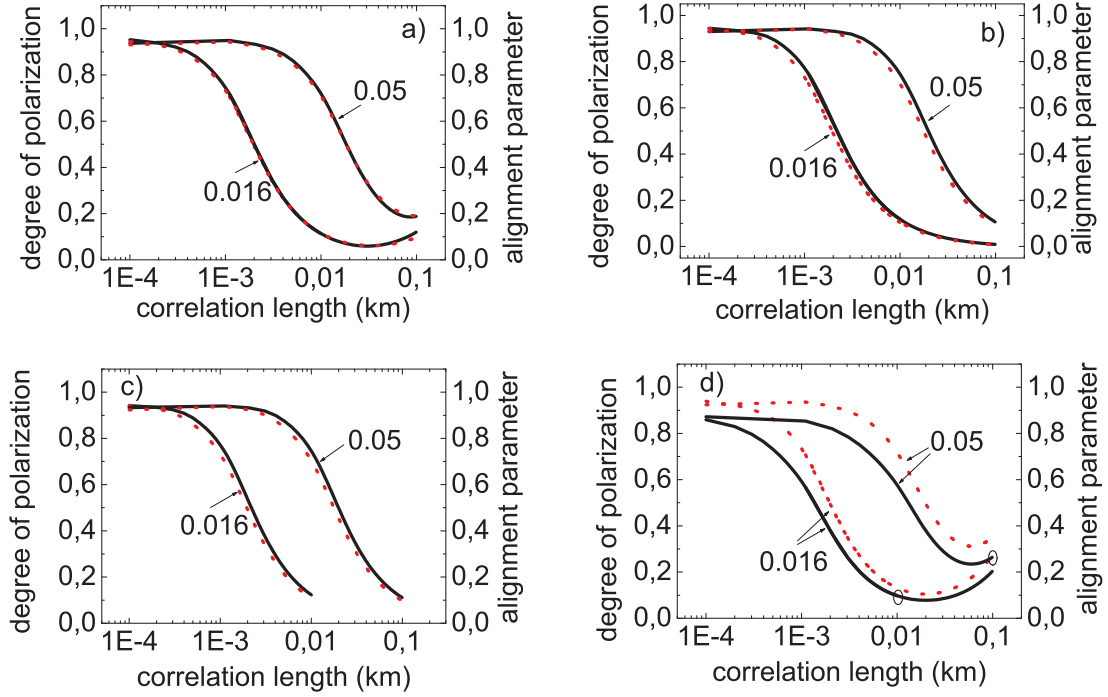}
\end{center}
\caption{DOP of the signal beam (black, solid) and
alignment parameter $A_{\uparrow\downarrow}$
(red, dotted) as function of correlation length $L_c$
for four pump SOPs: a) $3^{-1/2}(1,\, 1,\, 1)$;
b) $(1,\, 0,\, 0)$; c) $(0,\, 1,\, 0)$; d) $(0,\, 0,\, 1)$.
The value of the birefringence length
$L_B(\omega_p)$ is indicated on the plots in km.
Two circles on plot d) indicate one (of infinitely
many) pair of points with equal PMD coefficients.
Other parameters are: input signal power $0.1$~mW,
input pump power $8$~W, Raman gain
$g_0=0.6$~(W$\cdot$km)$^{-1}$, nonlinearity
parameter $\gamma =1$~(W$\cdot$km)$^{-1}$,
linear fiber losses $\alpha =0.2$~dB/km, total length
of the fiber $L=2$~km.}
\label{ris2}
\end{figure}
%=============

We start with the characterization of the performance
of the Raman polarizer in the counter-propagating
geometry by following the $D$ and
$A_{\uparrow\downarrow}$ for four different SOPs of
the pump beam and two values of the beat length, as a
function of the correlation length. The corresponding results are shown
in the four panels of Fig.~\ref{ris2}. All four plots
demonstrate that small correlation lengths and large
beat lengths favor the efficiency of Raman polarizers.
In the first three plots the maximum DOP is $0.94$,
while on the fourth plot it is of only $0.84$. Such large difference
in the maximal achievable DOPs gives a hint that the
performance of Raman polarizers is rather sensitive to the
choice of the pump SOP.

%=== Figure 3 ===
\begin{figure}
\begin{center}
\includegraphics[scale=0.6]{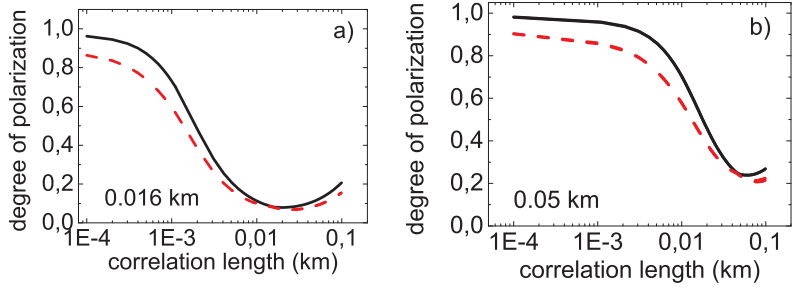}
\end{center}
\caption{DOP of the signal beam for two pump
SOPs as function of the correlation length $L_c$
for the value of the birefringence length
a) $L_B=0.016$~km and b) $L_B=0.05$~km. Other
parameters are the same as in Fig.~\ref{ris2}.
These two pump SOPs are those which
maximize (black, solid) and minimize (red,
dashed) the DOP of the output signal beam.}
\label{ris3}
\end{figure}
%=============

How sensitive is this dependence is demonstrated
by the two panels of Fig.~\ref{ris3}. For generating each
point on these plots we took $256$ realizations of the
pump SOPs, and for each of this SOP we followed the signal
DOP as a function of the correlation length. We plot only
those two realizations which provide maximum and
minimum DOP. From the analysis of these plots
we may conclude that from the viewpoint of efficient Raman
polarizers only a limited range of pump SOPs leads to a
good performance, while the rest of pump SOPs perform poorly. The
particular SOP which maximizes the DOP depends on
both the correlation and the beat length values.

\begin{figure}
\begin{center}
\includegraphics[scale=0.6]{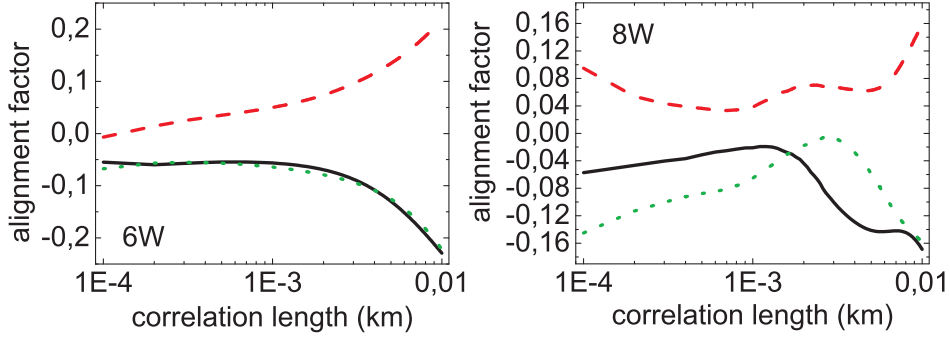}
\end{center}
\caption{Alignment factors defined in
Eqs.~(\ref{AF1})-(\ref{AF3}):
$A_{\uparrow\downarrow}^{(1)}$ (black solid);
$A_{\uparrow\downarrow}^{(2)}$ (red dashed);
$A_{\uparrow\downarrow}^{(3)}$ (green dotted),
as function of the correlation length for $L_B=0.05$~km
and for two values of the pump power:
a) $6$~W; b) $8$~W.}
\label{ris4}
\end{figure}

In all of the four plots of Fig.~\ref{ris2} we may observe that
the alignment parameter is relatively close to unity
whenever the DOP is more than $0.9$. From this result
we draw the conclusion that on average the SOP
of the outcoming signal beam is related to the
pump SOP as dictated by the alignment parameter that was
introduced in Eq.~(\ref{36}). A more detailed information
on the signal SOP is available when basing ourselves on the following definitions of the
alignment factors
\begin{eqnarray}
&& A_{\uparrow\downarrow}^{(1)}=
\left\langle
\frac{S^{(s)}_1}{S_0^{(s)}}
-\frac{S^{(p)}_1}{S_0^{(p)}}
\right\rangle\, ,
\label{AF1}\\
&& A_{\uparrow\downarrow}^{(2)}=
\left\langle
\frac{S^{(s)}_2}{S_0^{(s)}}
+\frac{S^{(p)}_2}{S_0^{(p)}}
\right\rangle\, ,
\label{AF2}\\
&& A_{\uparrow\downarrow}^{(3)}=
\left\langle
\frac{S^{(s)}_3}{S_0^{(s)}}
-\frac{S^{(p)}_3}{S_0^{(p)}}
\right\rangle\, ,
\label{AF3}
\end{eqnarray}
which are calculated at the Raman polarizer output $z=L$. These expressions quantify the pairwise
proximity of the components of the signal and
pump Stokes vectors. The dependence of the
alignment factors upon the correlation length for
$L_B=50$~m and for two values of the pump
power is plotted in Fig.~\ref{ris4}. One may
observe that the output signal SOP depends on
the pump power. Therefore, signal polarization
stabilization may be deteriorated by relative intensity
noise of the pump laser source.

%=== Figure 4 ===
%=============

Another important issue is the selection of fibers
for Raman polarizers. Usually, this selection relies
on the value of the PMD coefficient. In most practical
situations, this coefficient fully characterizes a randomly
birefringent fiber. However, in some cases the knowledge
of this parameter only is not sufficient for making a
conclusion on the performance of Raman polarizers, a fact that
was first noticed in Ref.~\cite{short}. For example the plot
in Fig.~\ref{ris2}(d) demonstrates that two fibers with
equal PMD coefficients lead to
Raman polarizers with rather different performances: in one
case $D\approx 0.26$, while in the other case it is $0.1$ only. 
In the regime of ideal Raman polarizers
(with DOP$>0.9$) and in the considered parameter range,
the PMD coefficient is indeed a reliable characteristic, in the
sense that fibers with equal PMD coefficients provide
Raman polarizers with similar performances. Nevertheless,
it is often desirable to consider the beat length and the
correlation length separately, as we do in this study,
rather than unite them under the single PMD coefficient.
Note that for $L\gg L_c$, which is always the case for
Raman polarizers, the PMD coefficient is defined as in
Ref.~\cite{wai_menyuk},
\begin{equation}
D_p=2\sqrt{2}\pi\frac{\sqrt{L_c}}{L_B}\omega_s^{-1}\, .
\label{37}
\end{equation}

%=== Figure 5 ===
\begin{figure}
\begin{center}
\includegraphics[scale=1.0]{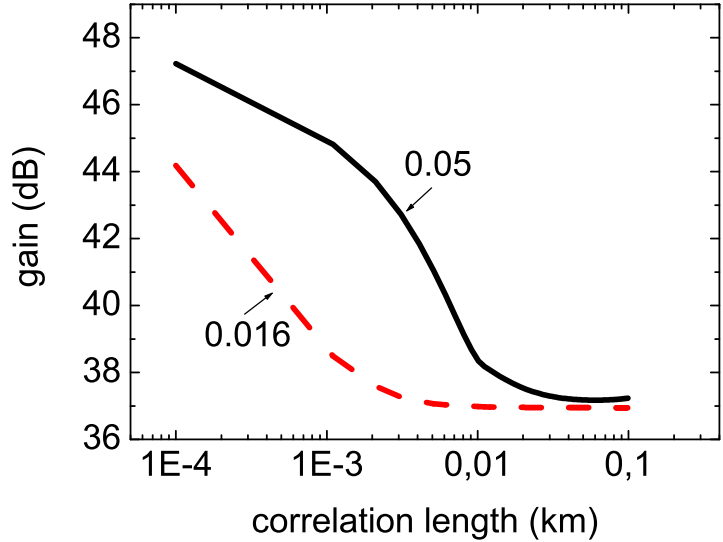}
\end{center}
\caption{Gain of a Raman polarizer as function of
the correlation length. The state of polarization of
the pump beam is $(1,\, 0,\, 0)$, the signal beam is
unpolarized. Values of the beat length $L_B$ are
indicated on the plot in km. Other parameters are
the same as in Fig.~\ref{ris2}.}
\label{ris5}
\end{figure}
%=============

Finally, let us consider the gain characteristics of
Raman polarizers. Owing to their strong
polarization-dependent gain, the gain of a
Raman polarizer is sizably larger than that of typical
polarization insensitive Raman amplifiers
(i.e., with high PMD values or that use
polarization scrambled pump beams), see
Fig.~\ref{ris5}. This increase in gain demonstrates
that Raman polarizers are simultaneously very
efficient Raman amplifiers. This feature means
that the amplification of signal beams with powers in
the mW range will almost certainly lead to the
depletion of the pump. This regime is left aside
in this study, as our goal was the demonstration of
the basic properties of Raman polarizers only. The regime
with depleted pump is so rich and parameter-sensitive
that it becomes rather difficult to draw general conclusions.
However, if necessary our theory is capable of treating
the depleted regime too.

%===============
\section{Conclusions}
We have developed a general theory for describing the interaction
of two beams in randomly birefringent fibers via the Kerr and
Raman effects. This theory can be applied to both the
co-propagation and counter-propagation configurations.
Here we considered only the counter-propagating geometry,
while details on the co-propagating regime can be found in
Ref.~\cite{short}. A comparison between these two studies
shows that the counter-propagating case is more demanding, in that it
requires for instance PMD coefficients below
$0.008$~ps/$\sqrt{\hbox{km}}$ for getting DOP slightly above
$0.9$ in a $2$~km fiber. Whereas the co-propagating geometry
can provide DOPs as high as $0.99$ for a $D_p$ as large as
$0.014$~ps/$\sqrt{\hbox{km}}$ with only a $1.5$~km long fiber.
In both cases the pump power is $8$~W.

However, it is important to point out that the counter-propagating configuration has the
advantage of clamping the signal SOP to the SOP of the
pump source. In the co-propagating geometry the output
signal SOP is also related to the pump SOP, but both
depend on the stochastically changing birefringence of
the fiber, thus complicating the control over the output
signal SOP.

In addition to the theoretical model of the Raman amplification
in randomly birefringent fibers, we presented the scheme for the
quantification of the performances of Raman polarizers. We
have identified three main characteristics of Raman polarizers: the
DOP of the outcoming signal beam, its SOP defined in relation
with the pump SOP, and the amplifier gain. The present study was limited
to the undepleted regime only, although the theory is readily
applicable to the depleted regime as well.

%=====================
\section{Acknowledgements}
We would like to thank L. Palmieri for valuable comments.
This work was carried out in the framework of the "Scientific
Research Project of Relevant National Interest" (PRIN 2008)
entitled "Nonlinear cross-polarization interactions in photonic
devices and systems" (POLARIZON), and appeared also
as a result of  2009 Italy-Spain integrated action "Nonlinear
Optical Systems and Devices" (HI2008-0075).

%====================

\end{document}